\documentclass[conference, 10pt]{IEEEtran}

\usepackage[utf8]{inputenc}
\usepackage[T1]{fontenc}
\usepackage[table]{xcolor}
\usepackage{soul}
\usepackage{amsmath}
\usepackage{amsfonts}
\usepackage{graphicx}
\usepackage{url}
\usepackage{cite}
\usepackage{color}
\usepackage{xspace}
\usepackage{booktabs}
\usepackage{makecell}
\usepackage{multirow}
\usepackage{rotating}
\usepackage{tikz}
\usepackage{calc}
\usepackage{booktabs}
\usepackage[hidelinks]{hyperref}
\usepackage[noend]{algpseudocode}
\usepackage{algorithm}
\usepackage{varwidth}
\usepackage[colorinlistoftodos]{todonotes}
\usepackage[caption=false]{subfig}

\usepackage{caption} 
\usepackage{amssymb}

\DeclareRobustCommand{\ffmd}{\emph{$F^2MD$}\xspace}
\DeclareRobustCommand{\fmeasure}{$\mathcal{F}_1$-measure\xspace}
\DeclareRobustCommand{\fname}{RADAR\xspace}
\DeclareRobustCommand{\ptf}{PTF\xspace}

\title{$RADAR$: a Radio-based Analytics for Dynamic Association and Recognition of pseudonyms in VANETs}

\author{%
    \IEEEauthorblockN{Giovanni Gambigliani Zoccoli, Filip Valgimigli, Dario Stabili, Mirco Marchetti}
    \IEEEauthorblockA{University of Modena and Reggio Emilia\\
    Department of Engineering ``Enzo Ferrari''\\
    \{giovanni.gambiglianizoccoli, filip.valgimigli, dario.stabili, mirco.marchetti\}@unimore.it
    } 
}

\begin{document}
\maketitle

\begin{abstract}
This paper presents \fname, a tracking algorithm for vehicles participating in Cooperative Intelligent Transportation Systems (C-ITS) that exploits multiple radio signals emitted by a modern vehicle to break privacy-preserving pseudonym schemes deployed in VANETs. 
This study shows that by combining Dedicated Short Range Communication (DSRC) and Wi-Fi probe request messages broadcast by the vehicle, it is possible to improve tracking over standard de-anonymization approaches that only leverage DSRC, especially in realistic scenarios where the attacker does not have full coverage of the entire vehicle path.
The experimental evaluation compares three different metrics for pseudonym and Wi-Fi probe \emph{identifier} association (\emph{Count}, \emph{Statistical RSSI}, and \emph{Pearson RSSI}), demonstrating that the \emph{Pearson RSSI} metric is better at tracking vehicles under pseudonym-changing schemes in all scenarios and against previous works. As an additional contribution to the state-of-the-art, we publicly release all implementations and simulation scenarios used in this work~\cite{zoccoli2025radarimplementation}.
\end{abstract}
\section{Introduction}
\label{s:intro}

In recent years, the rapid growth of vehicular technologies has led to the development of new solutions for improving road safety, traffic efficiency, and driver experience. The most promising solution is Vehicular Ad Hoc Networks (VANETs), a key technology for modern Cooperative Intelligent Transportation Systems (C-ITS)~\cite{Paul2017ITS}, which allow entities participating in the C-ITS to share their status and other data via the Dedicated Short Range Communication (DSRC) protocol~\cite{Toh2001AdHoc, 5888501}.

Despite their great potential for enabling future smart cities, VANETs require all entities to share large amouns of information that could be used to profile users, thus exposing them to targeted attacks. As an example, many VANETs applications require connected vehicles to share messages containing their precise location, thus allowing an adversary to track all the habits of their victims. 

To mitigate these risks, practitioners from both academia and industry have developed privacy-preserving mechanisms that are currently being adopted in VANETs communication. One of these solutions allows entities to hide their real identity by means of pseudonyms~\cite{petit2014pseudonym, gerlach2007privacy}, temporary identifiers that are used in VANETs messages to preserve the real identity of the participating entities. Several pseudonym change schemes have been proposed in the literature~\cite{etsipseudonyms}, however this approach is still not sufficient to prevent de-anonimization and tracking of vehicles.

In~\cite{zoccoli2023vanets}, the authors demonstrated the possibility to track the same vehicle despite the use of different pseudonym-changing schemes, by analyzing data such as position, speed, and acceleration with basic motion and trajectory formulas. 

\subsection{Motivations}
\label{ss:motivations}
While existing research on vehicle tracking in VANETs under pseudonym-changing schemes has demonstrated the feasibility of linking pseudonyms~\cite{benarous2020alloyed, zoccoli2023vanets}, all the available approaches rely on strong and often unrealistic assumptions about the attacker’s capabilities, typically requiring access to all VANET communications and a complete coverage of the full paths of tracked vehicles. 
We remark that these assumptions are not practical (especially in large-cities scenarios). Since the effectiveness of the attack is limited to the areas monitored by the attacker, this de-anonymization approach poses a limited threat for real-world deployments.

To address this gap, in this paper we propose \fname, a Radio-based Analytics for Dynamic Association and Recognition of pseudonyms in VANETs. \fname considers a more realistic threat model, where the adversary only has access to radio-based communications monitored with antennas that are placed on few non-overlapping areas.
\fname uses two primary sources of signals generated by a connected vehicle: DSRC (as already analyzed in~\cite{zoccoli2023vanets}) and Wi-Fi. 
Wi-Fi is often used in modern In-Vehicle Infotainment (IVI) systems and is commonly found on any personal device of the passengers of the vehicle.
Experimental evaluation against existing methodologies and different pseudonym-changing schemes demonstrates the effectiveness of \fname in tracking vehicles under pseudonym-changing schemes, urging the development of more efficient schemes to preserve the privacy of drivers in VANETs.

\subsection{Contributions}
\label{ss:contributions}
The contributions of this work to the state-of-the-art are threefold. First, we introduce a novel tracking methodology that enables an attacker to monitor vehicles participating in VANET communications by leveraging multiple radio-based technologies. This approach allows effective vehicle tracking even when the attacker does not have access to all transmitted messages, thereby exposing the limitations of current pseudonym-changing schemes in preserving user privacy.
Second, we conduct an experimental evaluation of three distinct tracking metrics, assessing their effectiveness against $5$ different pseudonym-changing schemes. We further compare the best-performing metric with existing methods, demonstrating that our approach outperforms prior work across various scenarios.
Third, we release the full implementation and simulation setups used in our evaluation~\cite{zoccoli2025radarimplementation}. To the best of our knowledge, this is the first publicly available implementation of a vehicle tracking methodology. We hope this contribution will serve as a foundation for future research and encourage further experimental advancements in this area.


\subsection{Manuscript Organization}
\label{ss:organization}

The rest of this paper is organized as follows. Section~\ref{s:related} reviews existing literature in the vehicle-tracking field, while Section~\ref{s:background} provides the required knowledge necessary for understanding of this work. Section~\ref{s:etf} introduces the design and implementation of \fname, discussing the supported metrics, while Section~\ref{s:evaluation} presents the experimental evaluation and comparison of our methodology with previous work. Finally, Section~\ref{s:conclusion} summarizes the contributions of this work.
\section{Related work}
\label{s:related}
Most related work focuses on presenting a novel pseudonym-changing scheme, with only a few references discussing the effectiveness of these schemes and their resilience against de-anonymization attacks.

The first significant analysis is presented in~\cite{8100875}, where the authors analyzed and classified several pseudonym-changing schemes into two main categories: \emph{mix-zone} and \emph{mix-context}. The former includes methodologies that use the vehicle’s location to change the pseudonym, while the latter involves approaches that decide whether a pseudonym change is necessary based on different input factors. In~\cite{8100875}, the authors also provided an analysis of selected pseudonym-changing schemes aimed at preserving driver privacy in a simulated environment. They highlighted that the only schemes harder to track in such an environment are those based on radio silence. Although these schemes are effective at hiding pseudonym changes in VANET communication by introducing a period of radio silence~\cite{benarous2020alloyed}, they have also been criticized for being incompatible with safety-related applications. As a result, many practitioners do not recommend their use in real-world scenarios~\cite{6737592}.

Another methodology evaluated as effective in preserving the privacy of VANET users is presented in~\cite{singh2019cpesp}. Specifically, the authors propose a \textit{Cooperative Pseudonym Exchange and Scheme Permutation (CPESP)}, a technique based on a dual approach to enhance privacy. This methodology can be applied in two distinct scenarios. In the first, CPESP requires a high number of vehicles and changes the pseudonyms of all participating entities simultaneously to reduce traceability. In the second scenario, CPESP operates in environments with fewer vehicles by randomly employing either a silence period or a periodic pseudonym-change scheme. Experimental evaluation against the same attack scenario described in~\cite{buttyan2009slow} demonstrates the effectiveness of this approach in protecting user privacy. However, all the aforementioned studies present only an experimental evaluation of their proposed pseudonym-changing schemes against a single malicious entity, without comparing their results to alternative tracking methodologies or providing a benchmark for evaluation.
Recent works, such as~\cite{huso2025frequency,irfan2024preventing}, focus on analyzing and mitigating the privacy risks associated with physical-layer signal characteristics demonstrating how Radio Frequency Fingerprinting (RFF) can be exploited by an attacker to track users based on the characteristics of the physical devices. The authors also introduce \emph{FingerJam}, a novel approach that employs controlled low-power jamming to obfuscate device-specific signal features, thereby obstructing RFF-based identification without compromising communication quality. While both approaches contribute valuable insights into physical-layer privacy preservation, they differ from our work, which focuses on analyzing message content and Received Signal Strength Indicator (RSSI) values rather than physical-layer signal characteristics.

The first paper presenting a benchmark comparison of the effectiveness of pseudonym-changing schemes against a malicious actor is presented in~\cite{zoccoli2023vanets}, where the authors presented an evaluation of the effectiveness of privacy-preserving schemes included in~\cite{F2MD} (the reference framework for security and privacy studies on VANET communication). The results presented in~\cite{zoccoli2023vanets} demonstrate that complex pseudonym-changing schemes do not provide more robust privacy guarantees, with an attacker with access to VANETs communication being able to track the same vehicle under different pseudonyms with \fmeasure higher than $0.91$.
However, the tracking framework presented in~\cite{zoccoli2023vanets} only considers an attacker accessing all VANET communication as its threat model, thus limiting the attacker tracking capabilities on the only area covered by its antenna.

To address this issue, in this work we propose a novel approach for tracking VANETs communication that considers a more realistic threat model where the attacker is able to monitor multiple radio-based communications with distinct and non-overlapping antennas over a wider area of the target city.
In particular, the methodology presented in this work exploits the DSRC protocol used in VANETs communication and the Wi-Fi probing messages generated by the In-Vehicle Infotainment system of a modern vehicle. 
To the best of our knowledge, this work is the first one presenting a methodology for dynamic association and recognition of pseudonyms based on multiple radio-based communication protocols.

\section{Background Knowledge}
\label{s:background}
In this section we present the basic knowledge required to understand the rest of this manuscript. In Section~\ref{ss:vantes}, we provide the details of VANETs communication and the current methodologies to ensure privacy of the communications; while in Section~\ref{ss:wifi}, we present the fundamentals of the Wi-Fi protocol that are exploited by \fname to track vehicles in VANETs.

\subsection{VANETs, DSRC and Pseudonyms}
\label{ss:vantes}
VANETs are dynamic networks that enable vehicles to exchange real-time data with each other and with infrastructure, aiming to improve road safety, traffic efficiency, and user experience. At their core lies C-ITS, which supports V2V and V2I communication through frequent data exchange (e.g., position, speed, road conditions) to enable cooperative and informed driving decisions.

Effective C-ITS deployment relies on standardized protocols to handle challenges like high mobility and dynamic topology. The primary standard is IEEE \emph{802.11p}, also known as WAVE, which operates in the $5.9$ GHz band with a $23$ dBm power limit ($200$ mW~~\cite{klapevz2021experimental, sae_j2945}) and enables low-latency communication via DSRC.

At the application layer, SAE J2735 defines the Basic Safety Message~\cite{J2735} (BSM), broadcast by vehicles every $100-–200$ ms. BSMs include mandatory state information (e.g., GPS, speed, braking) and optional data for enhanced functionality. Each BSM also includes an identifier field, which is expected to contain the digital identity of the entity sending the message to enable identification. However, many researchers expressed privacy concerns related to the usage of an identifier in a broadcast-type network, developing numerous strategies to reduce tracking risks and protect sensitive user information. 
One of the most promising solutions to this issue is represented by certificate pseudonyms, a set of pre-loaded, certified public keys stored in the vehicle~\cite{8100875}, which are used instead of a static identifier. While the usage of pseudonyms is supported by multiple VANETs standards (such as IEEE~\cite{IEEE1609.2} and ETSI~\cite{ETSI_TS_102_941_V2_2_1}), there is no a standard procedure describing how pseudonyms should change, thus leading to a variety of pseudonym-changing schemes being proposed by researchers.

Basic pseudonym-changing strategies are often based on parameters such as elapsed time, number of messages sent, or distance traveled~\cite{F2MD}. More advanced approaches use contextual factors (such as the presence of nearby vehicles) to trigger changes~\cite{F2MD, gerlach2007privacy, pan2013cooperative}. Despite the variance of the proposed schemes presented in literature, there are still two major issues that are not covered by previous research. First, several proposals require modifications to the standard BSM structure~\cite{pan2013cooperative, liao2009effectively, Boualouache2020, wang2018trigger}, making them non-compliant with the IEEE 1609 standard. The second and most critical issue is the lack of evaluation of pseudonym-changing schemes against realistic adversaries. In fact, while many works only present a novel pseudonym-changing scheme to demonstrate its feasibility from a computation perspective, only a couple of existing works have discussed the privacy-preserving effectiveness of existing methodologies~\cite{singh2019cpesp, zoccoli2023vanets}, despite using an unrealistic and impractical adversarial model. 

\subsection{Wi-Fi}
\label{ss:wifi}
The Wi-Fi protocol standard, defined in the IEEE $802.11$ standard~\cite{9363693}, is a wireless communication protocol enabling high-speed data transfer across short to medium distances. Operating in unlicensed frequency bands, primarily $2.4$ GHz and $5$ GHz, Wi-Fi employs techniques such as \emph{Orthogonal Frequency Division Multiplexing} (OFDM) to achieve efficient and reliable data transmission. The Wi-Fi protocol supports various modulation schemes and channel widths, allowing for flexible deployment across a range of environments and applications, enabling internet access, media streaming, IoT connectivity, and device-to-device communication.

Modern IVI systems offer Wi-Fi communication to the vehicle's passengers, as either a simple hotspot for sharing cellular connectivity or for more complex applications. Hence, it is now common to find integrated cellular modems in the Telecommunication Units (TCUs) of a modern IVI system, which offer the same features found in a typical wireless access point. 
As described in the Wi-Fi $IEEE~802.11$ protocol~\cite{9363693}, when a personal device wants to identify the list of nearby available Wi-Fi networks, it sends a broadcast Wi-Fi probe request. Upon reception of the probe request, the IVI system (if the hotspot feature is enabled) responds with a probe response frame, containing all data required by the client to connect to the Wi-Fi access point. This probe response contains $34$ different fields, with the most important being: 
\begin{itemize}
    \item \texttt{Service Set Identifier (SSID)}: the name of the network as set by the access point;
    \item \texttt{Media Access Control (MAC)}: the physical address of the device broadcasting the Wi-Fi probe. 
\end{itemize}

To prevent loss of probe response, each access point is usually configured to send these messages with a default period of $100~ms$, with a maximum transmission power equal to $20~dBm$ (equivalent to $100~mW$). 
\section{Design and implementation of \fname}
\label{s:etf}
In this section we describe the design and the implementation of \fname. Our tracking algorithm relies on three metrics for linking the messages sent by the vehicles to the unique \emph{identifier} exposed by the Wi-Fi access point in Section~\ref{ss:wifi}.
Our threat model considers an attacker with a limited number of antennas to passively monitor VANET communication (i.e., DSRC messages) and Wi-Fi probes received in proximity to the antennas. We remark that this threat model is extremely cost-efficient, since by placing antennas strategically it is possible to track many vehicles even without covering their whole paths.

\begin{figure}[htb]
    \centering
    \includegraphics[width=\columnwidth]{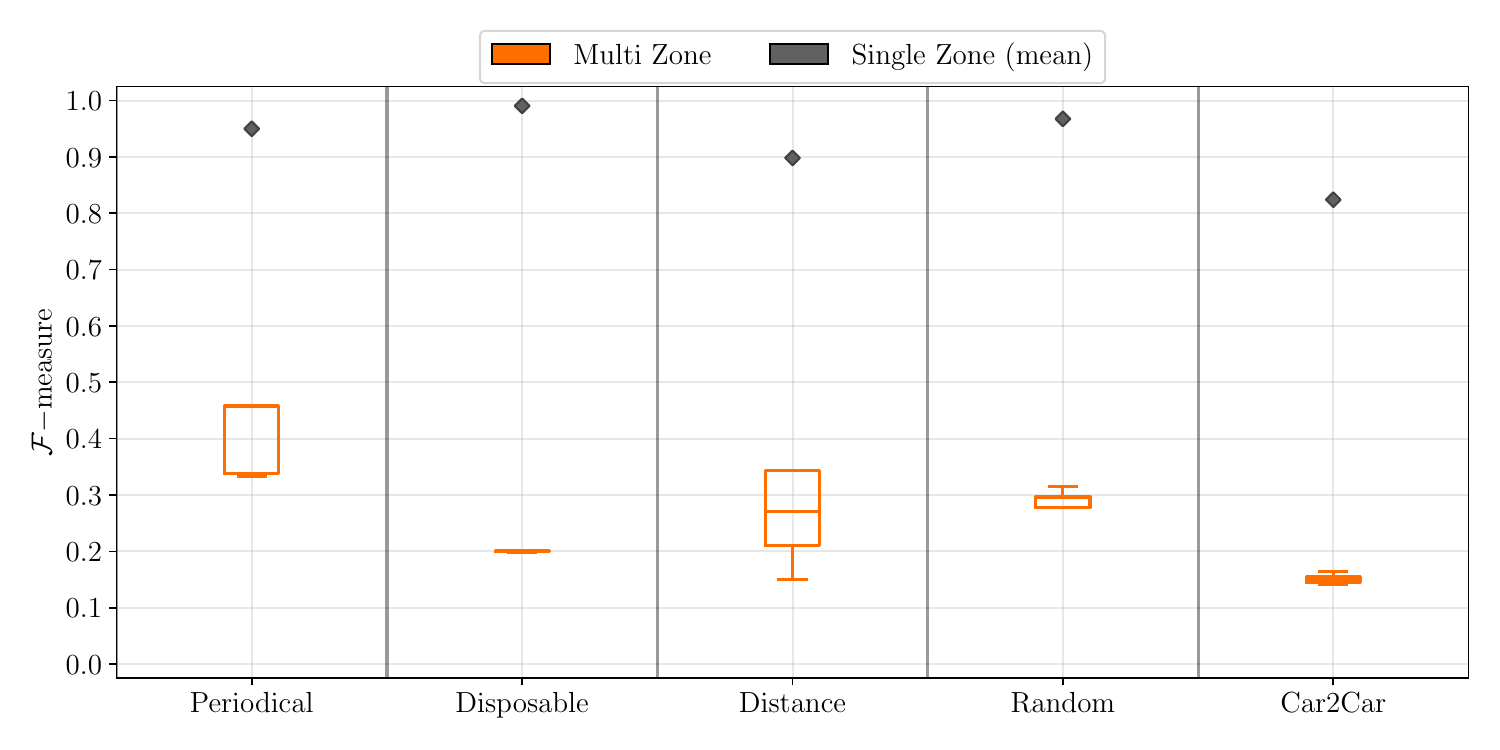}
    \caption{Multi-zone and single-zone performance comparison of \ptf against different pseudonym-changing schemes of \ffmd with $1~Hz$ of sending frequency }
\label{fig:PTF_single_multi}
\end{figure}

For tracking vehicles' pseudonyms we employ the same strategy presented in the Pseudonym Tracking Framework~\cite{zoccoli2023vanets} (\ptf), which is extremely effective in linking the different pseudonyms belonging to the same vehicle inside the coverage area of a single antenna, with peak performance of $0.9$ \fmeasure against multiple pseudonym-changing schemes. However, \ptf performance drops in a realistic scenario in which the attacker monitors multiple and non-overlapping areas of the map. The motivation is that \ptf cannot link vehicles whose pseudonym changes outside of the area covered by the attacker. Figure~\ref{fig:PTF_single_multi} shows the clear drop in performance from the ``single-zone'' scenario considered by the related work~\cite{zoccoli2023vanets} (gray points showing high \fmeasure) to the ``multi-zone'' scenario considered in this work (the \fmeasure over different runs is shown by the orange box-plots). The detailed explanation of the pseudonym-changing schemes used in the comparison presented in Figure~\ref{fig:PTF_single_multi} is available in the original manuscripts~\cite{zoccoli2023vanets, F2MD}.

As an effort to overcome this limitation, the tracking methodology developed in this work leverages Wi-Fi probe \emph{ID} of the vehicle's access point as an additional source of information to track a vehicle. 
We remark that 
this methodology can be extended to include any other wireless communication protocol (such as Bluetooth Address~\cite{sig:core_spec} or TPMS~\cite{grys2019experimental}). 

The analysis presented in \fname comprises three phases. In the first phase we use the same strategy presented in \ptf~\cite{zoccoli2023vanets} to identify all the different pseudonyms associated with the same vehicle inside a single coverage area. 

In the second phase we define a list of candidate Wi-Fi probe \emph{IDs} by selecting all the IDs that are received by the antenna within the same time frame as the pseudonyms associated to the same vehicle by \ptf. At the end of the second phase we associate a single Wi-Fi probe \emph{ID} to each vehicle within each coverage area. In this work we will evaluate and compare (in Section~\ref{s:evaluation}) the performance of three different heuristics:
\begin{itemize}
    \item \emph{Count}: considers the number of beacons received between the DSRC and the Wi-Fi probe; 
    \item \emph{Statistical RSSI}: considers a simple analysis of the signal strength of DSRC and Wi-Fi messages;
    \item \emph{Pearson RSSI}: considers a more sophisticated analysis of the signal strength of DSRC and Wi-Fi messages.
\end{itemize}

Finally, in the third phase of \fname we use the Wi-Fi probe \emph{ID} to reconstruct the trip of the vehicles across multiple non-overlapping areas. 

The full pseudo-code description of \fname is summarized in Algorithm~\ref{alg:tracking}.

\begin{algorithm}[htb]
    \caption{Vehicle tracking}
    \label{alg:tracking}
    \begin{algorithmic}[1]
        \Function{MatchVehicle}{$BSMs$}
            \State {$ P_{groups} = \Call{MatchPseudonyms}{BSMs} $} \label{line:ptf}
            \For{$ seq $ \textbf{in} $ P_{groups} $}
                \State {$ BMS_{msg} = \Call{GetMessages}{seq} $}
                \State $b_{match} = \Call{FindMatch} {\newline \phantom{spaces usefull to align} unique_{ids},~BSM_{msg},~metric}$ \label{line:match}
                \State {$ \Call{LinkVehicle}{b_{match},~BSM_{msg}} $}
            \EndFor
        \EndFunction
    \end{algorithmic}
\end{algorithm}

The details of the three different metrics are presented in the next sections.

\subsection{Count metric}
\label{ss:count}
The \emph{Count} metric considers only the number of messages collected by the attacker for selecting the best Wi-Fi probe \emph{ID}. This metric considers the number of DSRC messages composing the list of pseudonyms associated with the same vehicle by \ptf and the number of beacons of the Wi-Fi probe with the same \emph{ID}. This metric assumes that vehicles do not change the probing frequency during operation. 

\subsection{Statistical RSSI metric}
\label{ss:rssi}
The \emph{Statistical RSSI} metric uses the signal strength of the different messages (DSRC and Wi-Fi beacons) for the correlation of the pseudonyms and a unique Wi-Fi probe \emph{ID}. The selection of the best Wi-Fi probe \emph{ID} is based on the weighted average of the differences between $7$ statistical indexes evaluated over the RSSI of the DSRC messages and Wi-Fi beacons. 
The selection of the Wi-Fi probe \emph{ID} is based on the minimum value of the average differences (i.e., the most similar RSSI to the reference values of DSRC messages) with different weights associated for every index. The list of indexes and their corresponding weights are presented in the following:
\begin{itemize}
    \item \textbf{mean} RSSI (value): $0.1$
    \item \textbf{standard deviation} RSSI (value): $0.3$
    \item \textbf{median} RSSI (value): $0.1$
    \item \textbf{max} RSSI (value): $0.05$
    \item \textbf{min} RSSI (value): $0.05$
    \item \textbf{max} RSSI (timestamp): $0.2$
    \item \textbf{min} RSSI (timestamp): $0.2$
\end{itemize}

We remark that these weights are selected following an experimental validation phase to maximizes the performance of this metric.
The detailed description of the RSSI metric is presented in Algorithm~\ref{alg:rssi_metric}.

\begin{algorithm}[htb]
    \caption{RSSI metric}
    \label{alg:rssi_metric}
    \begin{algorithmic}[1]
        \Function{FindMatch}{$unique_{ids},~BSMs,~RSSI$}
            \State {$ BSM_{rssi} = \Call{GetRSSI}{BSMs} $}
            \State {$ mean_{dsrc} = \Call{mean}{BSM_{rssi}} $}
            \State {$ std_{dsrc} = \Call{std}{BSM_{rssi}} $}
            \State {$ median_{dsrc} = \Call{median}{BSM_{rssi}} $}
            \State {$ min_{dsrc} = \Call{min}{BSM_{rssi}} $}
            \State {$ max_{dsrc} = \Call{max}{BSM_{rssi}} $}
            \State {$ mints_{dsrc} = \Call{mints}{BSM_{rssi}} $}
            \State {$ maxts_{dsrc} = \Call{maxts}{BSM_{rssi}} $}
            
            \For{$ u $ \textbf{in} $ unique_{ids} $}
                \State {$ b = \Call{GetMessages}{u} $}
                \State {$ statistics[u].append(mean_{dsrc} - \Call{mean}{b}) $}
                \State {$ statistics[u].append(std_{dsrc} - \Call{std}{b}) $}
                \State {$ statistics[u].append(median_{dsrc} - \Call{median}{b}) $}
                \State {$ statistics[u].append(min_{dsrc} - \Call{min}{b}) $}
                \State {$ statistics[u].append(max_{dsrc} - \Call{max}{b}) $}
                \State {$ statistics[u].append(mints_{dsrc} - \Call{mints}{b}) $}
                \State {$ statistics[u].append(maxts_{dsrc} - \Call{maxts}{b}) $}

                \State {$ statistics[u] = \Call{AvgWeights}{\newline \phantom{used to align the function} statistics[u],~weights} $}
            \EndFor
            \State {$ b_{match} = \Call{GetMinimum}{statistics} $} \label{line:bmatch}
            \State \Return $b_{match}$
        \EndFunction
    \end{algorithmic}
\end{algorithm}

\subsection{Pearson RSSI metric}
\label{ss:complex_rssi}
The \emph{Pearson RSSI} metric uses a more sophisticated approach based on the \emph{Pearson} correlation~\cite{sedgwick2012pearson} for the selection of the Wi-Fi probe \emph{ID} corresponding to the reference pseudonyms.
We use the correlation between the pure RSSI values collected for the DSRC messages and the Wi-Fi beacon, and associate a list of pseudonyms with the Wi-Fi probe \emph{ID} exhibiting the highest value of Pearson coefficient. We remark that a preliminary step of interpolation is required to align the length of samples in the two sequences. The detailed description of the \emph{Pearson RSSI} metric is presented in Algorithm~\ref{alg:improved_rssi_metric}.

\begin{algorithm}[htb]
    \caption{Improved RSSI metric}
    \label{alg:improved_rssi_metric}
    \begin{algorithmic}[1]
        \Function{FindMatch}{$unique_{ids},~BSMs,~RSSI$}
            \State {$ DSRC_{rssi} = \Call{GetRSSI}{BSMs} $}
            \State {$ DSRC_{rssi} = \Call{Interpolate}{DSRC_{rssi}} $}
            \For{$ u $ \textbf{in} $ unique_{ids} $}
                \State {$ u_{rssi} = \Call{GetRSSI}{u} $}
                \State {$ u_{rssi} = \Call{Interpolate}{u_{rssi}} $}
                \State{ $ pearson[u].append(\Call{Pearson}{u_{rssi},~DSRC_{rssi}}) $ } \label{line:pearson}
            \EndFor
            \State {$ b_{match} = \Call{GetBestMatch}{pearson} $}
            \State \Return $b_{match}$
        \EndFunction
    \end{algorithmic}
\end{algorithm}
\section{Performance evaluation}
\label{s:evaluation}
In this section, we present the performance evaluation of \fname. In Section~\ref{ss:scenario} we present the simulation scenario used in our evaluation, providing all the required details to replicate our experiments. Then, in Section~\ref{ss:vtrace_results} we compare the tracking performance of the three different metrics of \fname. Finally, in Section~\ref{ss:previous_work_comp} we compare the results of the best-performing metric of \fname against previous work.
 
\subsection{Simulation scenario}
\label{ss:scenario}
\begin{figure}
    \centering
    \includegraphics[width=.98\columnwidth]{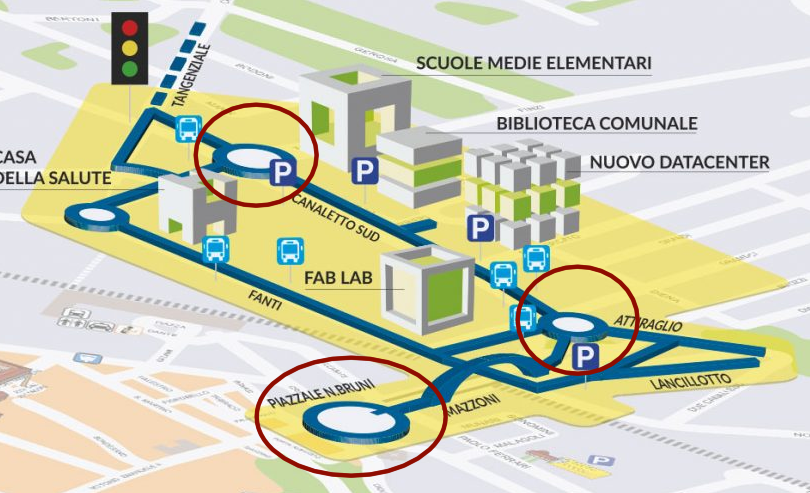}
    \caption{Graphical representation of the area covered by the attacker (red circles) inside the MASA}
    \label{fig:masa_antennas}
\end{figure}

The simulation scenario used to demonstrate the effectiveness of \fname is based on a modified version of the \ffmd framework~\cite{F2MD}, a widely used platform for misbehavior detection in VANETs under various scenarios. To simulate a realistic environment, we extended the framework to support the Wi-Fi probing protocol discussed in Section~\ref{ss:wifi}.
The experimental setup is based on the VEINS simulator~\cite{Veins}, an open-source framework for simulating vehicular communication networks, built on \emph{OMNeT++}~\cite{OMNET} and \emph{SUMO}~\cite{SUMO2018}. We simulated the \emph{Modena Automotive Smart Area} (MASA) with a total of 500 vehicles over 20 minutes of simulation, where the attacker had deployed three antennas at three intersections, as shown in Figure~\ref{fig:masa_antennas}~\cite{masa}.

To replicate the realistic behavior of real communication networks, we increased the \emph{transmission power} from $100~mW$ to $200~mW$, as explained in~\cite{klapevz2021experimental}. We also enabled the VEINS obstacle shadowing model to capture the impact of large buildings and other obstructions on signal transmission~\cite{6544519, 5720204}.

Since BSMs can be sent with different time intervals depending on the scenario, we tested the tracking performance of \fname in the worst-case scenario where the lowest number of messages is received, using a sending frequency of $1~Hz$.
We performed $25$ tests overall to remove any possible bias from the results. Moreover, to simulate a simple attacker with access to commonly available equipment, the radius of each monitoring antenna is set to $50~meters$, thus limiting the monitoring capabilities of the attacker.

Results are depicted by means of \fmeasure, expressing the harminc mean between the precision and the recall of the tracking capabilities of \fname, with values ranging from $0$ (no tracking at all) to $1$ (perfect tracking of all entities).

\subsection{Experimental Evaluation}
\label{ss:vtrace_results}
\begin{figure}[htbp]
    \centering
    \includegraphics[clip,width=.99\columnwidth]{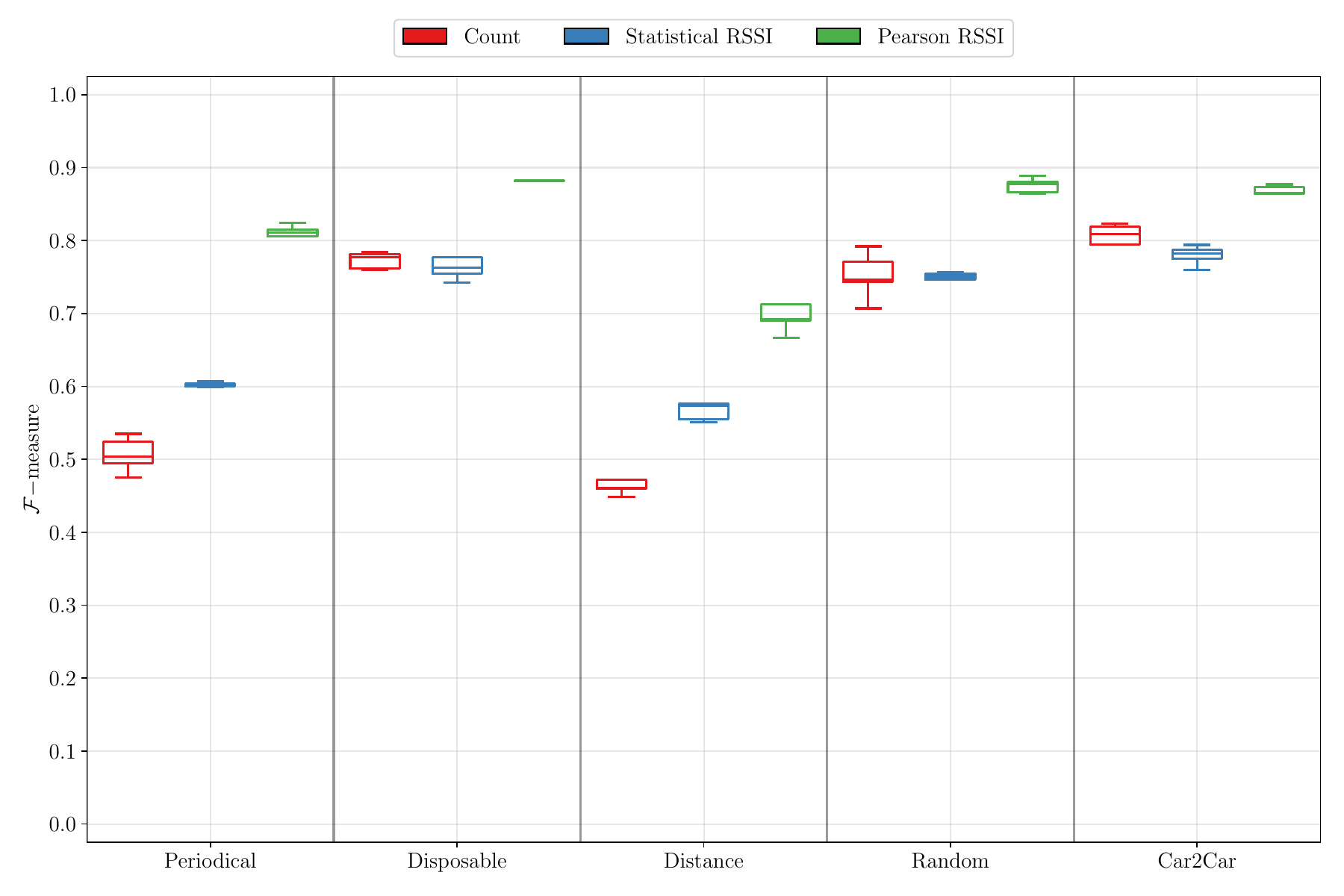}%
    \caption{Comparison of the $Count$, $Statistical~RSSI$ and $Pearson~RSSI$ metrics for tracking vehicles using a BSM sending frequency of $1~Hz$.}
    \label{fig:metrics_comparison}
\end{figure}

The experimental evaluation presented in this section focuses on the comparison of the proposed heuristics (see Section~\ref{s:etf}) against the different pseudonym-changing schemes supported in the \ffmd Framework~\cite{F2MD}. 
Figure~\ref{fig:metrics_comparison} shows the experimental results of \fname in the worst-case scenario, where BSMs are sent with a frequency of $1~Hz$. We compare the tracking performance of the \emph{Count} (Section~\ref{ss:count}) (depicted in red), the \emph{Statistical RSSI} (Section~\ref{ss:rssi}) (depicted in blue), and \emph{Pearson RSSI} (Section~\ref{ss:complex_rssi}) (depicted in green) metrics against $5$ different pseudonym-changing schemes (left-to-right): \emph{Periodical}, \emph{Disposable}, \emph{Distance}, \emph{Random}, and \emph{Car2Car}.
The results are depicted by using of box-plots to highlight the variance of the tracking performance between the different metrics.

The results presented in Figure~\ref{fig:metrics_comparison} clearly highlight that the \emph{Pearson RSSI} metric outperforms both \emph{Count} and \emph{Statistical RSSI} against all the pseudonym-changing schemes available in \ffmd. 
Overall, we highlight that the pseudonym-changing scheme with the lowest \fmeasure value is the \emph{Distance} scheme, where the pseudonym is changed based on the distance traveled by the vehicle.

\subsection{Comparison with previous works}
\label{ss:previous_work_comp}
In this section we present a two-fold comparison with previous work. We use the pseudonym-changing schemes presented in~\cite{singh2019cpesp} and the tracking methodology introduced in~\cite{buttyan2009slow} (named \emph{SLOWTrack}) to compare \fname against pseudonym-changing schemes not available in \ffmd and against a different pseudonym-tracking methodology. For the sake of completeness, we also evaluate the tracking performance of \emph{SLOWTrack}~\cite{buttyan2009slow} against the hardest pseudonym-changing scheme for \fname (\emph{Distance}).
We remark that there are no public implementations of either the pseudonym-changing schemes presented in~\cite{singh2019cpesp} or the \emph{SLOWTrack}~\cite{buttyan2009slow} tracking methodology. Hence, the results presented in this section are obtained by re-implementing both the schemes and the tracking algorithm based on the description available in the original works. 
Hence the comparison presented in this section has the two-fold objective of (i) comparing \fname with related work and (ii) serving as a benchmark for future advancements. As an additional contribution to the state-of-the-art, we publicly release all the implementations and simulation setups used in this work~\cite{zoccoli2025radarimplementation}.

\begin{figure}[htb]
    \centering
    \includegraphics[width=.98\columnwidth]{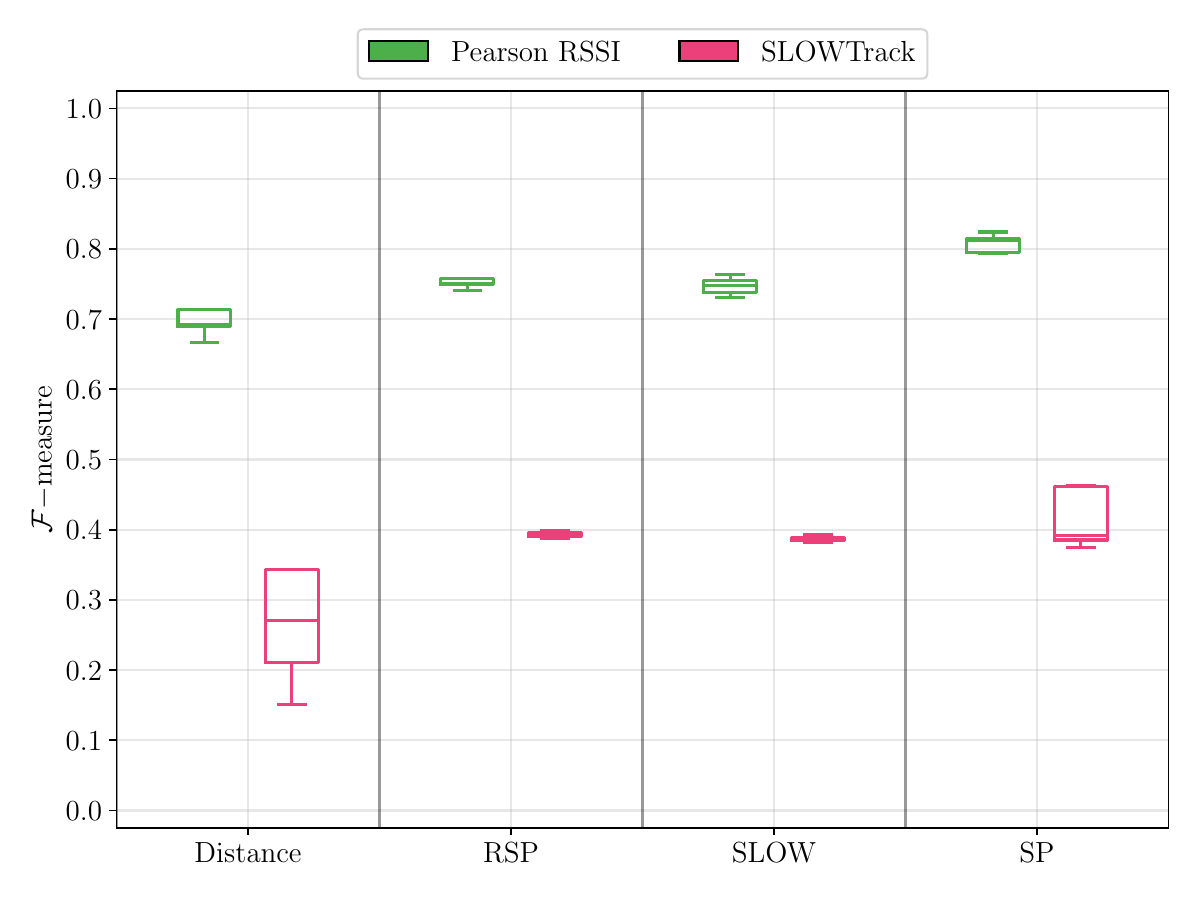}
    \caption{Experimental comparison of the \emph{Pearson RSSI} metric against the \emph{SLOWTrack} methodology over different pseudonym-changing schemes.}
\label{fig:comparison}
\end{figure}

The results of this comparison are summarized in Figure~\ref{fig:comparison}, where the \fmeasure achieved with \fname (\emph{Pearson RSSI} metric, green box-plot) is compared against \emph{SLOWTrack}~\cite{buttyan2009slow} (pink box-plot). 

From the analysis of the results presented in Figure~\ref{fig:comparison} it is clear that \fname outperforms \emph{SLOWTrack} in all simulated scenarios, reaching peak performance that is twice that of achieved by the other solution. 
The main cause of the low tracking performance of \emph{SLOWTrack} lies in the pseudonym-changing schemes used in our evaluation.  
In fact, all the pseudonym-changing schemes used in this comparison are based on a variable radio-silent period during which the pseudonym used by each vehicle is changed. However, in the original \emph{SLOWTrack}~\cite{buttyan2009slow} manuscript pseudonyms are tracked by ``\emph{joining the dots} between two heartbeat messages [...] or by constructing a trajectory through a consistent series of (position, velocity) pairs [...]''. This implies that, by using a radio-silent period in pseudonym-changing schemes, the tracking performance of \emph{SLOWTrack} is inversely proportional to the number of vehicles used in the simulation, while \fname is mostly unaffected. Hence, this demonstrates that our tracking methodology based on multiple radio sources is more resilient to radio-silence periods, thus effectively tracking vehicles in VANET communication.
\section{Conclusions}
\label{s:conclusion}

In this work we presented \fname, a tracking algorithm designed to exploit DSRC communication and Wi-Fi probe broadcast messages to bypass pseudonym-changing schemes and other privacy-preserving mechanisms in VANETs communication.
The experimental evaluation of \fname discusses three different metrics for tracking different pseudonyms belonging to the same vehicle, demonstrating that the metric based on the Pearson correlation between two series of RSSI signals (\emph{Pearson RSSI}) outperforms the other two metrics in the worst-case scenario, where messages are sent with a frequency of $1~Hz$, against $5$ different pseudonym-changing schemes. 
Experimental comparison against previous tracking algorithms and different pseudonym-changing schemes shows that \fname is able to achieve higher tracking performance in all scenarios, and is more resilient to radio-silent periods and missing data than previous work. We remark that these results highlight the need for stronger, multi-layered privacy-preserving solutions that account all the communications emitted from the vehicles and their passengers to effectively preserve the privacy of road users in VANETs communication.
As a final contribution, we also publicly release all the implementations and simulation setups used in this work~\cite{zoccoli2025radarimplementation} to overcome current limitations on reproducibility of previous work.

\section*{Acknowledgements}
This work has been supported by the project "FuSeCar" funded by the MUR Progetti di Ricerca di Rilevante Interesse Nazionale (PRIN) Bando 2022 - grant 2022W3EPEP.

This work has been supported by the Cyber Risks of Vehicle-to-Vehicle Communications (CRV2V) project (J33C22002810001) under the Cascade Open Calls of SERICS "CRV2V -- Cyber Risks of Vehicle-to-Vehicle Communications" (PE00000014) founded by MUR National Recovery and Resilience Plan.
\small{
    \bibliographystyle{IEEEtran}
    \bibliography{sections/bibliography}{}
}

\end{document}